# Gaze-informed Signatures of Trust and Collaboration in Human-Autonomy Teams

Anthony J. Ries[1,2], Stéphane Aroca-Ouellette[3], Alessandro Roncone[3], and Ewart J. de Visser[2]

1 DEVCOM Army Research Laboratory, Aberdeen Proving Ground, MD 21001 USA
2 Warfighter Effectiveness Research Center, U.S. Air Force Academy, CO 80840 USA
3 Department of Computer Science, University of Colorado Boulder, CO 80309 USA

**Abstract:**

In the evolving landscape of human-autonomy teaming (HAT), fostering effective collaboration and trust between human and autonomous agents is increasingly important. To explore this, we used the game Overcooked AI to create dynamic teaming scenarios featuring varying agent behaviors (clumsy, rigid, adaptive) and environmental complexities (low, medium, high). Our objectives were to assess the performance of adaptive AI agents designed with hierarchical reinforcement learning for better teamwork and measure eye tracking signals related to changes in trust and collaboration. The results indicate that the adaptive agent was more effective in managing teaming and creating an equitable task distribution across environments compared to the other agents. Working with the adaptive agent resulted in better coordination, reduced collisions, more balanced task contributions, and higher trust ratings. Reduced gaze allocation, across all agents, was associated with higher trust levels, while blink count, scanpath length, agent revisits and trust were predictive of the human's contribution to the team. Notably, fixation revisits on the agent increased with environmental complexity and decreased with agent versatility, offering a unique metric for measuring teammate performance monitoring. This is one of the first studies to use gaze metrics such as revisits, gaze allocation, and scanpath length to predict not only trust, but also human contribution to teaming behavior in a real-time task with cooperative agents. These findings underscore the importance of designing autonomous teammates that not only excel in task performance but also enhance teamwork by being more predictable and reducing the cognitive load on human team members. Additionally, this study highlights the potential of eye-tracking as an unobtrusive measure for evaluating and improving human-autonomy teams, suggesting eye gaze could be used by agents to dynamically adapt their behaviors.

## 1. Introduction

*1.1. Challenges in Human-Autonomy Teaming*

In the rapidly developing research of human-autonomy relationships, there is a heightened emphasis on fostering cooperation over competition within this evolving dynamic (Chiou & Lee, 2023; de Visser et al., 2020; O'Neill et al., 2022). This shift underscores an increasing acknowledgment of the synergistic potential that can arise through human-autonomy collaboration (Metcalfe et al., 2021). For example, manufacturing envisions future industrial settings where humans and robots, or cobots, work side by side (e.g. Weiss et al., 2021). Simultaneously, research in artificial intelligence aims to develop autonomous systems that augment human capabilities rather than replace them (Raisamo et al., 2019) and for good reason (Parasuraman et al., 2008; Parasuraman & Riley, 1997). Human-autonomy teams possess unique characteristics separate from human-human, human-automation, and human-animal teams (Lum & Phillips, 2024; McNeese et al., 2023; O'Neill et al., 2022; Phillips et al., 2018). Notably, humans may perceive their autonomous teammates differently and tend to express less trust in them (McNeese et al., 2021; Walliser et al., 2019;). This is particularly true when agents vary in their reliability (e.g. E. J. de Visser et al., 2016; E. de Visser & Parasuraman, 2011; Lu & Sarter, 2019; Walliser et al., 2023). Recent trust models suggest that achieving an optimal human-autonomy relationship will require a continuous mutually adaptive partnership (de Visser et al., 2018; de Visser et al., 2023). The human relationship with autonomous agents is transitioning from tool to teammate whereby each team member depends on the other to execute a task in support of an end goal (Groom & Nass, 2007; Klein et al., 2004; Lyons et al., 2021; Phillips et al., 2011; Walliser et al., 2023).

Recent calls have articulated the need to prioritize the design and development of AI teammates that understand and contribute to teamwork (Mangin et al., 2022; McNeese et al., 2023; Stowers et al., 2021). This means that while AI teammates are competent in executing task work, they must also be able to support teamwork such as goal alignment, fluent coordination, and social factors such as trust (E. J. de Visser et al., 2018; Kaplan et al., 2023; Wildman et al., 2024). This requires AI teammates to be developed and designed with the ability to anticipate the need of other teammates, be it that they are human or other AIs. To date, the focus has been to create AI that can reach the performance of top human players such as Go (Silver et al., 2016) and Starcraft (Vinyals et al., 2019). A different design objective is to collaborate closely together with humans to beat a game, requiring alternative approaches Challenges have emerged where groups of AI bots do not necessarily work well with their human counterparts in a multi-player real-time strategy game (Shah & Carroll, 2019). Human-Autonomy teams are therefore a unique challenge and should be seen as such (McNeese et al., 2023). There is thus a great need to investigate such teamwork functions that form a crucial interface between humans and autonomous agents. A critical question therefore is: How do we improve human-autonomy teaming? We propose two novel methods for improving human-autonomy teaming.

*1.2. Designing and Developing Adaptive AI Agents*

The first method to improve HAT is to create agents that are better able to adapt to their human counterparts. Early work on cockpit automation revealed the occurrence of "clumsy automation", where systems inadvertently increased workload during periods of high task demand (Wiener, 1989). Likewise, achieving true interdependence in HATs has been challenging due to the limited capabilities, the unreliability, and the misallocation of roles of automated agents resulting in the human teammate having to adapt to the AI agent to facilitate true teamwork . While the production of joint action is the hallmark of HAT interdependence, the intricacies of how and when specific subtasks are performed can be ambiguous. A proposed solution is the creation of *adaptive automation* that can anticipate the needs of the human user to balance mental workload, increase situation awareness and calibrate trust (E. A. Byrne & Parasuraman, 1996; Feigh et al., 2012; Parasuraman et al., 2009; Parasuraman & Wickens, 2008; Scerbo, 2018). Development of new types of agile agents promises to make the HAT co-adaptive where the machine agent adapts to the human agent and vice versa.

Unlike competitive tasks where the autonomous agent learns the optimal strategy, collaborative tasks often require both agents to have compatible, collaborative and predictable strategies to succeed. If an agent chooses to play a complex, but near optimal strategy with a human teammate that is new to the task, the team is likely to perform worse than if the agent exhibited a simpler more understandable strategy. Vice versa, if an agent cannot recognize human team player initiatives, team performance will suffer because opportunities for collaboration and better strategies are missed. The faster the human can infer the legibility of the agent's intended action, the quicker they can adapt to the agent's desired outcome (Dragan et al., 2013). Such legibility increases transparency of automated agent actions and enhances human-agent collaboration without necessarily escalating the overall human workload (Roncone et al., 2017; Stowers et al., 2021). As a result, traditional methods for training competitive agents such as self-play (SP) perform poorly with real humans in cooperative tasks since they are trained to perform the optimal but not necessarily the most human-compatible strategy.

Previous work has attempted to overcome this limitation by training an imitation learning agent to be a proxy human model, and then training a reinforcement learning (RL) model with the human proxy as the teammate (Carroll et al., 2019). This has provided some performance benefits, but this arrangement still drastically underperformed compared to human-human teams in most settings. More recent RL models using a hierarchical structure, such as the Hierarchical Ad Hoc Agent ($HA^2$), outperforms these other models and, importantly, is perceived by participants to be a more cooperative and intelligent team player (Aroca-Ouellette et al., 2023). Hierarchical reinforcement learning (HRL) agents, like humans, break down a complex problem space into manageable subtasks. An agent that learns a hierarchy of subtasks and how each subtask contributes to the overall task goal enables a more adaptive agent. This type of agent more closely aligns to the human at the behavioral and cognitive levels (Aroca-Ouellette et al., 2023). We hypothesize that more adaptive algorithms, like $HA^2$, will result in improved human-autonomy teaming processes, such as efficient performance monitoring and shared task load.

*1.3. Assessment of Human-Autonomy Teaming Factors with Eye Tracking*

Leveraging opportunistic, implicit human signals, such as eye movements, is the second method to enhance human-autonomy teaming. Autonomous agent models trained with human behavior often use keyboard and joystick inputs mapped to environment state changes (Carroll et al., 2019; Javdani et al., 2015). While useful, these inputs may be supplemented to provide a more detailed and holistic understanding of the human-agent relationship. Eye movements provide cost-efficient, unobtrusive measurements for predicting state changes in the human and offer a near real-time data source for agent adaptation (Marathe et al., 2018; Metcalfe et al., 2021; Neubauer et al., 2020). Notably, recent work shows that models combining user input with eye gaze information outperform baseline models in predicting human visual attention, proficiency, trust, and intent (Bera et al., 2021; Hulle et al., 2024; Thakur et al., 2023).

Human eye gaze produces multiple metrics for analyzing information selection and changes in cognitive state. Fixation revisits, which measure how often a person backtracks to an item, have revealed significant semantic and contextual differences in reading and search behavior (M. D. Byrne et al., 1999; Rayner, 1998). The proportion of eye gaze allocated to items can differentiate experts from novices and indicate periods of confusion (Law et al., 2004; Wachowiak et al., 2022). Additionally, scanpath length is related to visual processing strategies, with larger scan paths reflecting more global/ambient processing or planning using information outside the local field of view (Groner et al., 1984; Velichkovsky et al., 2002). Blinking behavior, influenced by task demands and fatigue, indicates that lower blink rates may reflect a strategy to maximize visual processing time or increased task engagement (Ranti et al., 2020; Veltman & Gaillard, 1996) while increased blink duration has been associated with higher fatigue, more errors, and reduced visual intake (Holmqvist et al., 2011; Morris & Miller, 1996). Gaze-derived features such as gaze location, sequence, and duration, and pupil diameter have informed image classification for computer vision (Karessli et al., 2017). Metrics like fixation frequency, average fixation duration, and pupil diameter predict cognitive workload fluctuations (Enders et al., 2021; Mallick et al., 2016; Pomplun & Sunkara, 2019; Van Orden et al., 2000, 2001; Wright et al., 2014).

Many eye movement features have proven highly informative in various teaming contexts. In a shared manipulation task, gaze patterns alone provided 83% classification accuracy between periods of robot automation and human teleoperation (Aronson et al., 2018; see also Aronson & Admoni, 2022). Similarly, eye gaze predicted users' intent in a sandwich-making task, with the most recently fixated item, fixation duration and frequency on items being the most predictive of subsequent selections (Huang et al., 2015). Additionally, incorporating eye data into a computational antagonist model resulted in fewer human wins, suggesting the model could infer and adapt to human strategies (Wetzel et al., 2014).

Eye gaze data also effectively predict changes in trust within automated systems (E. J. de Visser, Phillips, et al., 2023; Kohn et al., 2021; Krausman et al., 2022; Tenhundfeld et al., 2019). For example, Lu & Sarter, 2019 demonstrated that low reliability automation had lower trust ratings, increased fixation frequency, larger scan paths, and led to longer total fixation time compared to high reliability automation. A meta-analysis revealed a negative correlation between performance-based trust in automation and dwell time percentage (Sato et al., 2023), supported by a recent analysis of 13 studies, attributing 18% of the variance in automation monitoring via eye gaze to trust (Patton & Wickens, 2024). Comparable results were found in automated driving where increased trust in automation led to fewer fixations on the driving scene (Hergeth et al., 2016). More recent evidence shows that failures in highly reliable automation decrease trust and lead to more complex scan patterns as indexed with gaze entropy (Foroughi et al., 2023).

*1.4. Addressing the Research Gaps*

While prior research supports using gaze-derived metrics to complement human inputs for informing human-autonomy teaming behavior, much of this work has focused on human-human interactions, low levels of agent automation, or competitive agent teammates. Fewer studies have examined how multiple eye movement metrics can predict both trust and teaming behavior in cooperative teaming scenarios where humans collaborate with highly autonomous agents in fast-paced, variable environments to complete multiple subtasks. It remains unclear if gaze metrics are informative of human-agent teaming behaviors in these contexts and which metrics are most useful. For instance, eye movement measurements such as revisits, a common metric used in visual search and reading literature, may relate to the teaming construct of mutual performance monitoring, where teammates track each other's performance and errors (Salas et al., 2005). Additionally, tracking how gaze is allocated to different items in the environment could indicate team adaptability in situations where members must adjust their strategy based on the demands of their teammates or environmental context. Our study addresses this gap by combining an interactive teaming task with predictive modeling of gaze behavior, demonstrating how unobtrusive gaze signals can inform trust and teaming dynamics.

*1.5. The Current Study*

To evaluate our methods for improving human-agent teaming, we used the game Overcooked AI (Carroll et al., 2019) to measure behavior, trust, and various eye metrics in a cooperative task with varying agent models and environmental complexity. Our behavioral analyses focused on team scores, human-agent collisions, number of tasks completed and each member's contribution to the task. Post-trial ratings provided subjective measures of trust, team fluency and coordination. Saccade, fixation, and gaze measures in relation to agent behavior, environment complexity, team performance, and trust outcomes were also assessed. Two hypotheses guided our work. We predicted that performance and eye-tracking-related measures would be influenced by agent versatility and the navigational complexity of the environment (H1). We also predicted that eye tracking metrics would predict trust and teaming behavior (H2, see Table 1). The results of this study show the unique influence of agent behavior and navigational challenges on human-agent dynamics emphasizing the potential for improved collaboration. Specifically, our findings indicate that adaptive agents significantly enhance HAT performance and trust and eye gaze metrics effectively predict these outcomes.

We introduce and evaluate novel gaze-derived metrics such as fixation revisits to the agent and scanpath length per second, which have not been previously applied to the context of adaptive HAT. These metrics revealed nuanced relationships between environmental complexity, agent behavior, and human contribution to the team. While increased fixations on an unreliable agent may be expected, our finding that revisits decreased with agent versatility but increased with navigational complexity provides a more detailed behavioral signature of teammate monitoring. Furthermore, our predictive models demonstrate that simple gaze features such as reduced agent fixations and shorter scanpaths can indicate higher trust, supporting their utility for adaptive AI development.

## 2. Material and methods

*2.1. Participants*

A total of 83 volunteers from the Location 1, and Location 2 participated in the experiment. Participants were recruited through newsletter announcements and through the Sona online experimental recruiting software. Three participants were discarded due to technical difficulties with the system and 12 were removed due to poor eye tracking data quality (see Data Processing – Saccades, Fixations and Blinks below and Supplement A.1). As a result, data from 68 participants were used in all analyses. This final sample included 31 participants from Location 1 (Age: $M = 24.1$ years, $SD = 8$; Gender: 20 female, 9 male, 1-non-binary/third gender, 1 preferred not to respond) and 37 participants from Location 2 (Age: $M=19.2$ years, $SD = 1.9$; Gender: 20 male, 17 female).Before participating, volunteers signed an informed consent document approved by the institutional review board at Location 2 in accordance with the Declaration of Helsinki.  Participants at Location 1 were reimbursed with a $25 gift card while Location 2 participants received extra course credit. All participants had normal vision (20/40 or better) without contact lenses. Three participants wore eyeglasses.

*2.2. Environment for Studying Human/AI Cooperation*

Our experiment used Overcooked AI, an established paradigm for studying human/AI cooperation (Aroca-Ouellette et al., 2023; Carroll et al., 2019; Long et al., 2024; Strouse et al., 2021; Wachowiak et al., 2022). The Overcooked AI platform provides a high-fidelity abstraction of real-world human-autonomy teaming scenarios by requiring humans and agents to perform interdependent tasks under time pressure, manage shared spatial constraints, and coordinate in real time. These characteristics parallel challenges in collaborative environments such as disaster response, medical robotics, and military operations. Its widespread adoption as a benchmarking environment (Aroca-Ouellette et al., 2025; Carroll et al., 2019; Liang et al., 2024; Lou et al., 2023; Strouse et al., 2021) reflects its value in the field for studying teaming behavior like coordination, adaptability, and trust in a repeatable and systematically controllable framework. While simplified compared to real-world systems, Overcooked AI allows precise control of components of the environment, such as the individual and coordination complexity require to succeed, making it an effective platform for isolating the impact of agent behavior and environmental complexity on team performance.

This game requires a coordinated strategy between chefs (one human and one autonomous agent) to earn point rewards in a kitchen by performing a sequence of actions to cook and serve soup (see Figure 1). Different environmental layouts were used to induce varying levels of team interdependence and navigational complexity, while the training policies of the autonomous teammate changed their interactive teaming behavior with the human. To play the game, participants took three onions, one at a time, from the onion dispenser, placed them in a pot (see Figure 1), served the resulting soup on a dish and delivered it to the serving area as many times as possible within an 80-second time limit. Each soup served increased the game score by 20 points.

*2.3. Experimental Design*

The study design was a 3x3 repeated measures with navigational challenge (low, medium, high) and agent behavior (clumsy, rigid, adaptive) as the within-subjects variables. The navigational challenge varied based on previously used Overcooked AI environments (Aroca-Ouellette et al., 2023; Carroll et al., 2019). The environments were constructed so that the task goal of cooking and delivering soup could be done independently, but the coordinated navigation strategy and potential for teammate collisions differed.

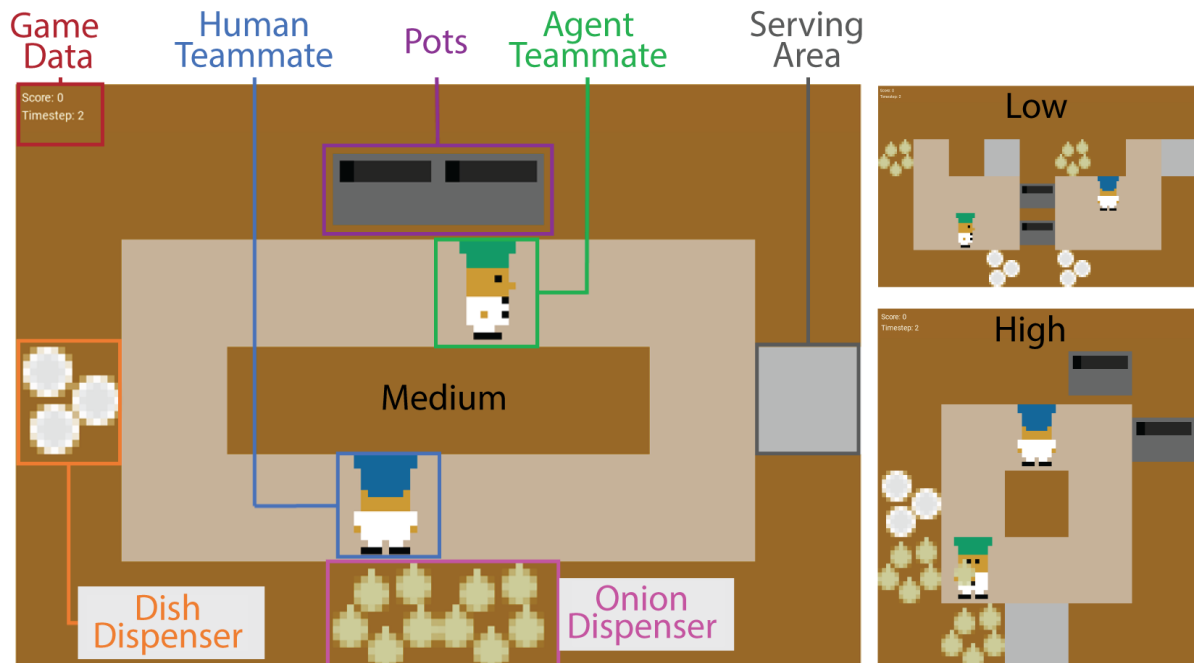

*Figure 1. Overcooked environments. The human and agent must work collaboratively to place three onions in a pot, wait for it to cook, place the cooked soup on a dish, and serve the soup. Each soup served earns a point reward. Left: Medium navigational challenge. Top right: Low navigational challenge. Bottom right: High navigational challenge*

In the low navigational challenge environment, known originally as "Asymmetric Advantages", the design ensures that the human and agent do not occupy the same space, making collisions impossible. The medium navigational challenge environment, originally known as "Counter Circuit", allows for potential collisions, but using the middle counter space to pass ingredients can mitigate this risk. In the high navigational challenge condition, also known as "Coordination Ring", navigation is difficult due to the smaller workspace and the ring-like design, which limits navigation to clock- or counter-clockwise movement. Each environment constituted a grid of tiles, with each tile measuring 180 x 180 pixels, subtending approximately 4 degrees of visual angle. The low navigational challenge environment was a 9x5 grid, the medium challenge was 8x5 and the high challenge was 5x5.

The agent behavior variable was based on previous work using these agents (Aroca-Ouellette et al., 2023). The clumsy agent selects actions at random, resulting in unpredictable and generally unhelpful behavior (Cook et al., 1991; Lee & Seppelt, 2012; Woods, 1996). The rigid agent, trained using self-play reinforcement learning, performs adequately in the environment but cannot adapt well to novel situations, making its behavior inflexible (Silver et al., 2017, 2018). In contrast, the adaptive agent, known as $HA^2$, was trained using hierarchical reinforcement learning and features human-inspired architecture with components for task selection and execution (Aroca-Ouellette et al., 2023). The adaptive agent was tuned for cooperative strategies, and designed for better adaptation to novel situations, new environments and different human teammates. HA$^2$ represents the current state of the art in cooperative human-agent teaming within the Overcooked AI framework. Unlike traditional self-play or imitation learning agents, HA$^2$'s explicit hierarchical task structure mirrors human planning strategies, making agents more understandable, enabling more fluent coordination, and improving overall team performance.

*2.4. Hardware and Software Resources*

The game was played on a standard desktop monitor (1920x1080 pixels; 53x30 cm) using a Dell Precision 7760 laptop computer (11th Gen Intel i7 2.3 GHz processor, 32 GB RAM, Windows 10 Pro). Participants used the keyboard arrow keys to control the directional movements of their game player and the space bar to interact with the game environment.

Eye tracking data were acquired from the Tobii Pro Spectrum (firmware 2.6.0) at 300 Hz and included the x,y position (converted to computer pixel coordinates), sample validity, pupil diameter, and eye openness (distance from upper to lower eyelid) from each eye. The ambient lighting in each recording location consisted of overhead fluorescent lights with no windows.

Gameplay ran at 5 fps, with each frame timestamped, recorded, and synchronized with other data streams using Lab Streaming Layer (LSL) and Lab Recorder (Kothe et al., 2024). LSL is a framework for synchronizing time-series data from multiple sources and sampling rates, providing a language-agnostic, real-time communication protocol for data recording and sharing. Game state information from each frame contained the x,y location in computer pixel coordinates of all items (e.g. pots, onions, agent, and human), the status of the game environment (e.g. if an onion was in a pot, soup cooking), agent behavior, navigation complexity, trial number, timestamp, number of human/agent collisions, and game score. The score was only affected by the number of soups served, and a collision was logged when the human and agent attempted to occupy the same tile.

All computers, monitors, eye trackers, and software were of the same brand, specifications and versions between recording sites.

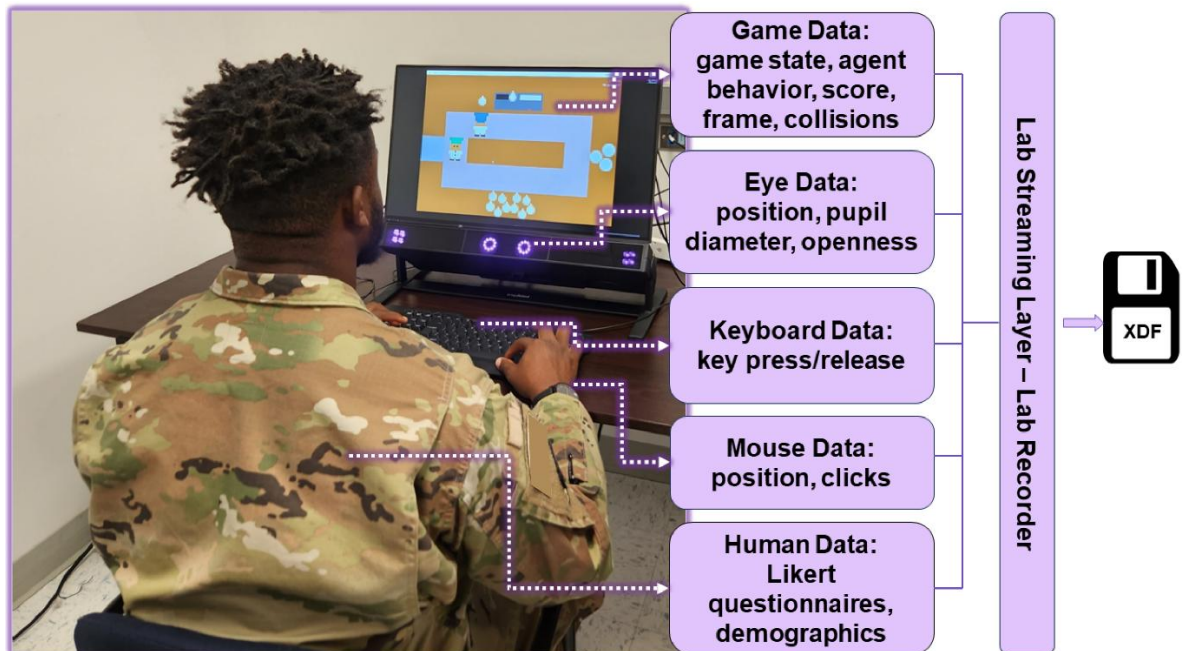

*Figure 2. Research environment. Frame-by-frame details of the game environment were synchronized with eye tracking and human behavioral inputs using Lab Streaming Layer and saved as an extensible data file (XDF).*

## 2.5. Procedure

After signing the informed consent, participants electronically completed a demographics form followed by a series of questionnaires, which are not analyzed in this work. A summary of prior gaming experience and Overcooked experience is shown in Supplement A.7. Participants were then seated approximately 70 cm in front of the computer monitor and eye tracker where a 5-point calibration was performed using the Tobii Eye Tracker Manager (2.6.0). Eye position was validated using the online gaze position, and recalibration was performed if, while looking at any calibration point, the gaze position exceeded 1.5 degrees. No participants failed the calibration procedure.

Next, data recording was initiated using the LSL Data Recorder to synchronize and store data streams from the eye tracker, mouse, keyboard, and game in one file (Figure 2). Participants were presented with task instructions on the computer monitor, allowed to ask questions, and given practice with the game. Practice continued until the participant served one soup and felt comfortable performing the task. The environment in the practice session was different from those in the experiment.

After practice, participants donned noise-cancelling headphones to prevent background distractions. No distractions were documented. Each participant completed two trials of each agent (clumsy, rigid, adaptive) and environment (low, medium, high navigational challenge) combination, totaling 18 trials. Trials were randomized and lasted 80 seconds each. Trials were self-paced and started when the participant pressed the Enter key. After each trial, participants rated five statements, adapted from Hoffman, 2019, on a 7-point Likert scale (1 = strongly disagree, 7 = strongly agree). Participants viewed the statements on the computer monitor and used the mouse to rate and submit their responses. The five statements consisted of 1) The human-agent team worked fluently together, 2) I was the most important team member, 3) I trusted the agent to do the right thing, 4) I understood what the agent was trying to accomplish, and 5) The agent was cooperative.

## 2.6. Data Processing

### 2.6.1. Saccades, Fixations, and Blinks

Before analysis, the proportion of missing data samples from the left pupil in each trial was calculated to assess eye data quality. To ensure high data quality while accommodating typical missing data from blinks or momentary tracking loss, we excluded participants if one or more trials had greater than 30% missing data. As a result, 12 out of the 83 participants were excluded from the study (Supplement A.1). For the remaining participants, averaged left and right x,y gaze positions were used to calculate saccades and fixations and averaged left and right pupil data were used for blink count and duration calculations. Prior to saccade, fixation and blink detection we linearly interpolated missing data segments consisting of up to 15 consecutive samples (50 ms).

Eye blinks were detected using the algorithm described in (Hershman et al., 2018), which calculates measurement noise that precedes and follows missing-sample time windows to determine blink onset and offset, i.e. blink duration. Saccades and fixations were detected using a velocity-based algorithm adapted from the EYE-EEG toolbox (Dimigen et al., 2011; Engbert & Mergenthaler, 2006). Thresholds were established using the median of the velocity time-series, smoothed over a 5-sample window. This method therefore calculates separate velocity thresholds for each participant. For all participants, we used a velocity factor of six (six times the median velocity), a minimum saccade duration of 12 ms and a minimum fixation duration of 50 ms.

To filter out biologically implausible saccades, a subsequent set of thresholds was applied[1]. Specifically, we removed saccades > 120 ms in duration and peak velocities less than 25 deg/sec or greater than 1000 deg/sec (Enders et al., 2021; Long et al., 2024). Microsaccades less than or equal to 1 deg in amplitude were not considered in the analysis. Fixations shorter than 75 ms or outside the monitor coordinates were also removed. Consistent with Holmqvist et al., 2011 we removed eye blinks with durations less than 75 ms or greater than 900 ms. After filtering these outlier events, we derived the metrics shown in Table 1 for each trial.

### 2.6.2. Gaze Mapping to Environment

The proportion of gaze samples allocated to game items was calculated for each trial. For every game frame, the location of each tile in screen coordinates, the tile label (e.g. onion, floor, counter) the current subtask being performed by the human and agent (e.g. getting onion from dispenser, serving soup), and the LSL timestamp was determined. Each eye gaze sample recorded during the game frame was initially labeled with the terrain type (e.g. counter, pot, serving station, dispenser) by co-registering the gaze sample position with the environment location label. If there was a game object (e.g. onion on counter) at the terrain type, the gaze label was overwritten with the game object type. If there was a player at the terrain type, the gaze sample was labeled as human or agent teammate. The soup object was created when an onion was placed in a pot or when the soup was on a dish before being served. The floor was any portion of the navigational path not containing a game object or player. If a gaze sample was recorded as not a number (NaN) (e.g. during a blink), or its position was outside the game-space monitor coordinates, it was labeled as 'other'.

---

[1] Velocity-based classifiers can sometimes misclassify eye events. For instance, when an eye blink occurs, the eye tracker will lose the pupil image. Upon recovery, it detects a significant change in the recorded eye position which is often mistakenly classified as a saccade since the duration and velocity exceed established thresholds. However, the recorded velocity is faster than the human eye can move and large in magnitude for a saccade of that duration.

*2.6.3. Interactive Team Behavior*

2.6.3.1. Subtasks

Interactive teaming behavior was assessed using the number of subtasks completed by each team member and their contribution toward the task goal of serving soup (Bishop et al., 2020; Long et al., 2024). Subtasks completed (e.g. get onion from dispenser, grab onion from counter, grab dish from dispenser, serve soup) was defined as the total number of subtasks completed within a trial, calculated separately for both the human and the agent. Team member contribution was defined as the difference between the relative proportion of human subtasks supporting the task goal and the relative proportion of the agent subtasks supporting the task goal. Team member contribution values ranged from 1 (all human contribution) to -1 (all agent contribution), with a contribution value of 0 indicating an equal contribution from both team members.

2.6.3.2. Revisits

We introduce revisits as a novel metric to evaluate human-agent teaming. Revisits are backtracking events within a scanpath sequence, typically described by the relationship between consecutive *saccades* where the saccade between fixation n+1 and fixation n+2 is in the opposite direction of the saccade between fixations n and n+1. We define revisits within a sequence of *fixations* where fixation n+2 returns to the same item as fixation n, and fixation n+1 is on a different item from n (Findlay & Brown, 2006). In other words, our definition of revisit occurs when an item is immediately refixated after fixating on a different item, regardless of saccade direction. Our analysis focuses specifically on agent revisits (e.g., fixate agent, fixate onion, fixate agent).

*Table 1. Definitions, descriptions, and predicted outcomes of eye metrics used in the analyses.*

| Eye Metric | Definition | Interpretation and Prediction | Related Constructs |
| --- | --- | --- | --- |
| **Scanpath Length** | Total of all saccade amplitudes (degrees of visual angle) divided by total trial duration (seconds) | Indicates how far the eye moves each second on average. Larger scanpath lengths may reflect more global as opposed to local processing or increased planning behavior to information outside local field of view. Prediction: Scanpath length is expected to increase with higher navigational challenge and less adaptive agent behavior due to more planning and monitoring of the agent/environment, resulting in decreased human task contribution. | Information Processing; Teammate Performance Monitoring; Team Coordination |
| **Fixation Duration** | Average duration (seconds) of a fixation | Represents the average time available for perceptual intake. May reflect different types of information processing. Prediction: Increased navigational challenge and less adaptive agent behavior will require more planning and monitoring of the agent/environment, resulting in shorter average fixation duration in these cases. | Information Processing; Teammate Performance Monitoring; Team Coordination |
| **Fixation Count** | Total number of fixations | Indicates spatial locations with increased information processing. Prediction: Increased navigational challenge and less adaptive agent behavior will require more planning and monitoring of the agent/environment, resulting in higher fixation counts. | Teammate Performance Monitoring; Team Coordination |
| **Revisits** | Number of times an item is immediately refixated after fixating on a different item (e.g. fixate agent, fixate onion, fixate agent) | Indicates how often the human immediately backtracks to an item. Increased revisits suggest more monitoring of and/or adapting to the revisited item. Prediction: Revisits will be negatively related to human task contribution; the more often the human revisits their teammate, the less task contribution they will make. | Teammate Performance Monitoring; Adaptability |
| **Gaze Proportion** | Number of eye samples having the same spatial location as an item/area, divided by the total number of samples, multiplied by 100 | Proportion of eye gaze allocated to an item or area. Indicates the relative importance, meaning, or trustworthiness of an item or area. Prediction: In line with human-automation research, increased eye gaze on the agent teammate will show an inverse relationship with trust. | Teammate Performance Monitoring; Trust |
| **Blink Count** | Total number of blinks | General indicator of processing demands and task engagement. Lower blink rates may reflect a strategy to maximize visual processing time and/or increased task engagement. Prediction: Higher engagement and more visual processing time, (i.e. lower blink count), will result in higher human task contribution. | Task Engagement; Processing Demands |
| **Blink Duration** | Average duration (seconds) of a blink | Average time from eye blink onset to offset. Increased blink duration is associated with higher levels of fatigue, increased errors, and reduced visual intake. Prediction: Human task contribution and blink duration will be inverstly related. | Fatigue; Errors |

Note. Metrics were individually calculated for each trial

## 3. Results

Behavioral and eye tracking analyses were performed in R (RStudio 2023.12.1+402 "Ocean Storm" Release). All dependent measures, except number of subtasks completed, were analyzed using a random effects repeated measures analysis of variance (ANOVA) with the 'ezANOVA' R package. The within-subjects factors were navigational challenge (low, medium, high) and agent teammate behavior (clumsy, rigid, adaptive) with participant as a random effect. The number of subtasks completed variable was analyzed using a mixed repeated measures ANOVA, with team member (human, agent) as the between-subject variable and navigational challenge (low, medium, high) and agent teammate behavior (clumsy, rigid, adaptive) as the within-subject factors. Greenhouse Geisser epsilon is reported with degrees of freedom corrections, and partial eta squared is reported for effect size. All complete ANOVA outputs have been provided in supplementary materials and are referenced in each section for efficient reporting. Detailed figures were produced to illustrate the relationship between the independent variables and the measures[2]. The grey non-overlapping whiskers (95% confidence interval) around the mean, marked by the green diamond, indicate significant differences between levels of each variable. T-tests were performed using the 'emmeans' and 'pairs' functions, while comparisons of teammate contribution against a zero value (equal contribution) were conducted using the 'contrast' function, all in R, with Bonferroni adjustments.

*3.1. Behavior*

3.1.2. *Collisions and Score*

Collision and score were analyzed separately, and each model showed significant main effects and interactions ($p < .05$, see Supplement A.2 for complete ANOVA output). As expected, there were more collisions in the high navigational challenge environment compared to the medium navigational environment ($t(67) = 10.03$, $p < 0.01$, Figure 3). No collisions were observed in the low navigational challenge environment because the human and AI teammate were 'physically' separated and could not collide. Across challenges, the rigid agent collided the most with the human, followed by the clumsy agent (rigid-clumsy = 6.75, $t(134) = 4.74$, $p < .001$). The adaptive agent had the lowest number of collisions (adaptive-rigid = -17.29, $t(134) = -12.15$, $p < .001$; adaptive-clumsy = -10.54, $t(134) = -7.41$, $p < .001$). This result indicates that the adaptive agent's behavior showed better coordination with the human compared to the other agents.

By far, the highest team scores were achieved in the low navigational challenge environment due to the convenient layout and the limited teamwork required to serve the soup efficiently. This allowed the human and AI teammate to complete tasks independently and quickly with occasional assistance from one another. This may also explain the lack of difference between rigid and adaptive agents in the low navigational challenge environment ($t(399) = 2.97$, p = 0.11), as the adaptive agent's ability to understand task hierarchy and anticipate the teammate's behavior was not advantageous when little teamwork was needed.

Despite the higher number of collisions, the high navigational challenge environment showed the second highest team score, ahead of the medium navigational challenge environment. The lower travel distance in this level may have contributed to more soups served as well as the adaptive agent's better navigation and coordination with the human in this environment.

---

[2]All box plots were created using the boxchart MATLAB (2021b) function. The horizontal line inside each box corresponds to the sample median while the top and bottom edges of the box are the upper and lower quartiles, respectively. Values that are more than 1.5 times the interquartile range (IQR) from the top or bottom are designated as outliers (points beyond whiskers). Each whisker above and below the box represents the maximum value that is not considered an outlier. The top and bottom edges of the notched region correspond to $m+(1.57 \cdot IQR)/\sqrt{n}$ and $m-(1.57 \cdot IQR)/\sqrt{n}$, respectively, where $m$ is the median, $IQR$ is the interquartile range, and $n$ is the number of data points, excluding NaN values. The green diamond represents the sample mean with gray error bars showing 95% confidence intervals. Significance between subgroups can be discerned by evaluating the 95% confidence intervals.

Based on scores, the medium navigational challenge environment proved to be the most challenging. Success in this environment required both navigational and strategic coordination by passing onions on the middle counter to reduce travel time. Again, the adaptive agent significantly outperformed the clumsy and rigid agents (adaptive-clumsy = 111.37, $t(134)$ = 47.37, $p < .001$; adaptive-rigid = 26.72, $t(134)$ = 11.36, $p < .001$) due to improved strategic and navigational coordination.

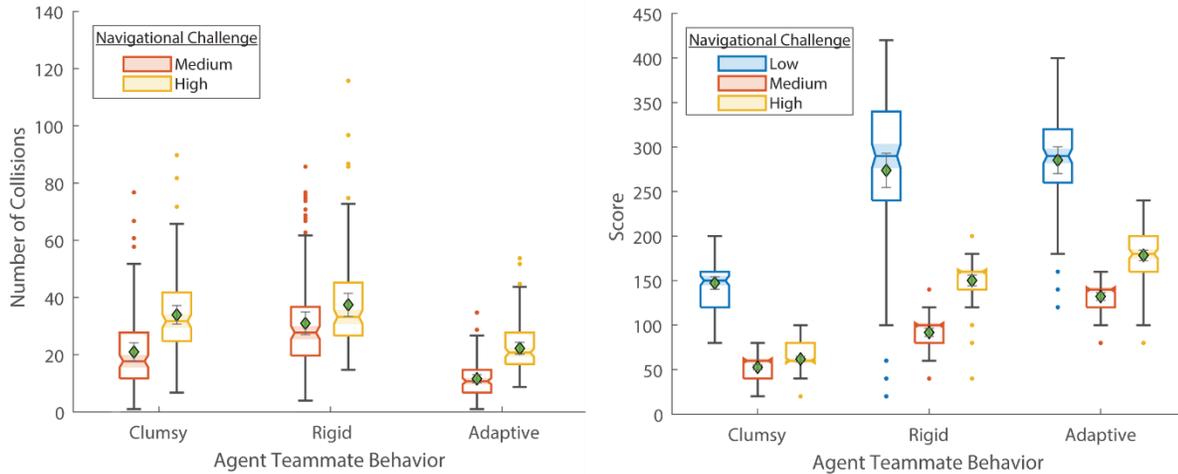

*Figure 3. Collisions (left) and score (right) for each agent, environment interaction. Note. Collisions were precluded in the low navigational challenge environment.*

### 3.1.3. *Interactive Team Behaviors*

Subtasks completed by each member and team member contribution were separately analyzed. Each model showed significant main effects and interactions ($p < .001$, Supplement A.3). Overall, the human completed significantly more subtasks than the agent, regardless of the teammate behavior (human-agent = 18.6, $t(134)$ = 46.4, $p < .001$) particularly in the low navigational challenge environment (low-medium =32.9, $t(134)$ =28.9, $p < .001$; low-high = 49.1, $t(134)$ = 43.1, $p < .001$) (Figure 4). On average, the adaptive agent completed significantly more subtasks compared to the other agents across environments (adaptive-rigid = 17.29, $t(134)$ = 26.0, $p < .001$; adaptive-clumsy = 42.6, $t(134)$ = 64.2, $p < .001$) and the rigid agent performed more subtasks than the clumsy agent (rigid-clumsy = 25.3, $t(134)$ = 38.1, $p < .001$). The significant agent/environment interaction showed the medium environment proved most challenging for both the rigid and adaptive agents, where they completed fewer subtasks compared to the low and high navigational challenge environments.

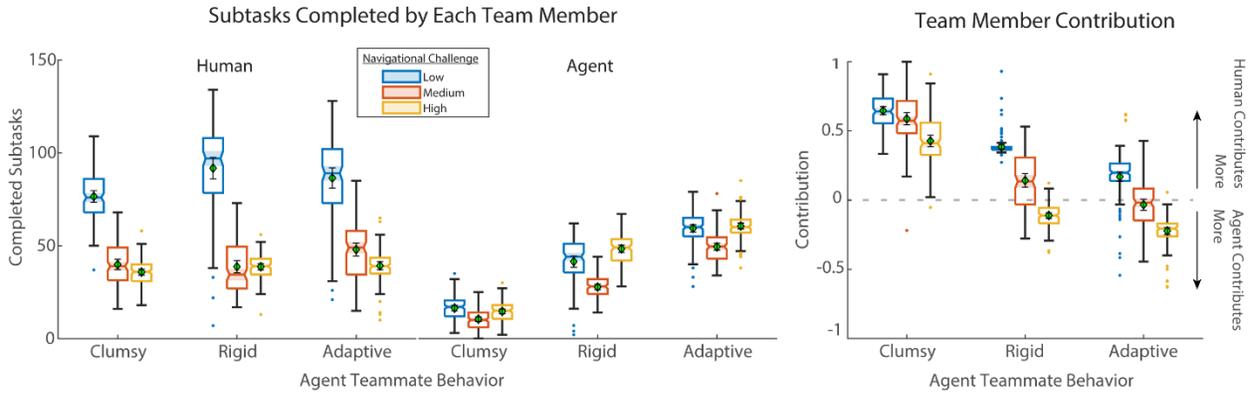

*Figure 4. Total subtasks completed (left) and contribution of each team member (right) for each agent, environment interaction. Right: The horizontal dashed line indicates equitable human, agent contribution.*

The human, on average, performed most of the work in each environment (low = .40, $t(144)$ = 38.7, $p < .001$, medium = .23, $t(144)$ = 22.4, $p < .001$, high = .03, $t(144)$ = 3.01, p = .009). The human contributed more when working with the clumsy (.55, $t(144)$ = 53.6, $p < .001$) and rigid (.14, $t(144)$ = 13.4, $p < .001$) agents. The adaptive agent reversed this dynamic with it contributing to teamwork overall more than the human (-.03, $t(144)$ = -2.8, p = .02). The significant interaction between navigational challenge and agent teammate behavior suggests that for the clumsy agent, the human was doing most of the work in the low navigational environment, followed by the medium and high environments. However, with the adaptive agent this pattern adjusted: the human still did most of the work in the low navigational challenge environment (.17, $t(396)$ = 11.6, $p < .001$) but shared the task load in the medium environment (-.03, $t(396)$ = -2.3, not significant, p = .2) while the agent contributed more when in the high navigational environment (-.22, $t(396)$ = -15.5, $p < .001$). These results demonstrate that the adaptive agent more effectively manages teaming and creates a more balanced task distribution across environments compared to the other autonomous agents.

### 3.2. Self-Report Measures

Each self-report statement was separately analyzed and showed significant main effects and interactions ($p < .001$, Supplement A.4). Overall, participants perceived the adaptive agent as the most cooperative (adaptive-clumsy = 4, $t(134)$ = 37.33, $p < .001$; adaptive-rigid = 0.65, $t(134)$ = 6, $p < .001$; rigid-clumsy = 3.4 $t(134)$ = 31.33, $p < .001$) and trustworthy (adaptive-clumsy = 3.97, $t(134)$ = 37.62, $p < .001$; adaptive-rigid = 0.56, $t(134)$ = 5.32, $p < .001$; rigid-clumsy = 3.4 $t(134)$ = 32.3, $p < .001$) (Figure 5). This pattern was most pronounced in the medium navigational challenge environments. Participants viewed themselves as the most important team member when working with the clumsy agent whereas this subjective perception was lowest when working with the adaptive agent, indicating a stronger teaming dynamic with the adaptive agent.

The highest fluency between the team was perceived with the adaptive agent followed by the rigid and clumsy agents. Fluency differences between agents, specifically the adaptive and rigid agents, were most significant in the medium navigation challenge environments ($t(397) = 6.44$, $p < .001$). Fluency differences between the rigid and adaptive agents disappeared in the high navigational environment ($t(397) = 0.97$, $p = 1$). This same pattern of results was displayed when participants rated their understanding of the agent's goal. When teaming with humans in environments of low navigational complexity, the rigid and adaptive agents were rated as more fluent, trusted, understandable, and cooperative compared to the other environments.

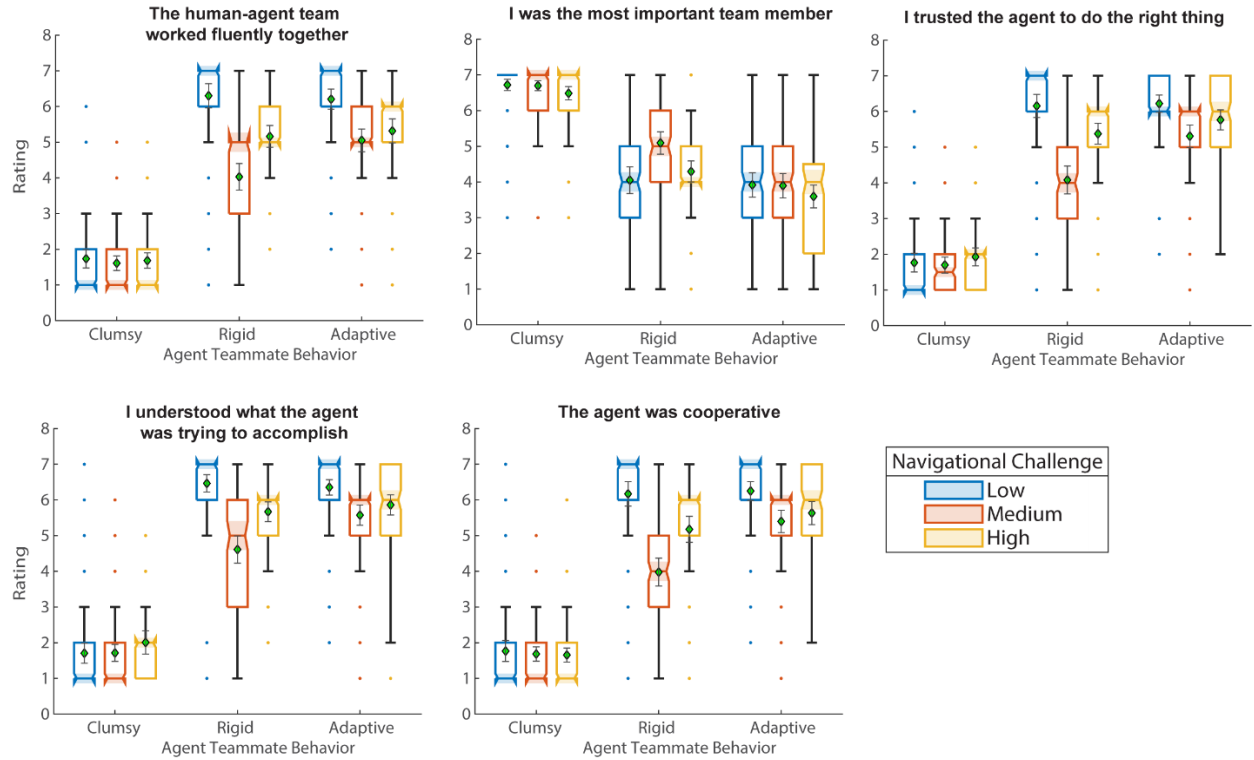

*Figure 5. Subjective agent ratings for each agent, environment interaction.*

### 3.3. Eye metrics

#### 3.3.1. Fixations and Saccades

Fixation count, fixation duration, scanpath length, and agent revisits were separately analyzed with each showing significant main effects and interactions ($p <= .001$, Supplement A.5). Eye movement summaries for saccades and fixations, collapsed across participants and agent behavior, are shown in Figure 6. Figure 6A shows the spatial distribution of fixation concentration (i.e. heatmaps) in each environment, which should be interpreted with caution as they do not depict human/agent locations. Figure 6B shows the distribution of fixation durations indicating generally longer fixations in the low navigational complexity environment relative to medium and high. The saccade main sequence in Figure 6C exemplifies the expected relationship between saccade amplitude and velocity (Bahill et al., 1975). The distribution of saccade angles in Figure 6D shows differences in saccade direction between environments.

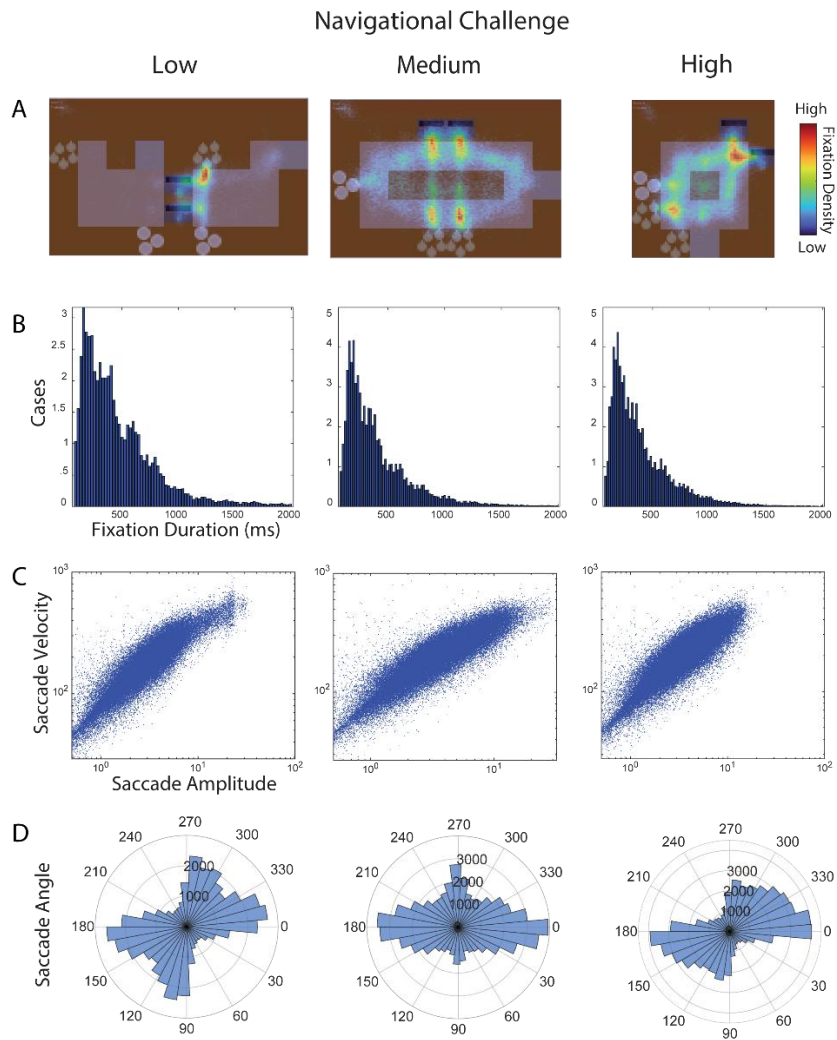

*Figure 6. Saccade and fixation summary across all participants for each navigational challenge. A. Heat maps indicate where most fixations were concentrated. B. Distributions of fixation duration. C. Main sequences showing the relationship between saccade amplitude and velocity. D. Saccade angular histograms. Outer ring number indicate angular degrees while the inner ring number represent fixation count.*

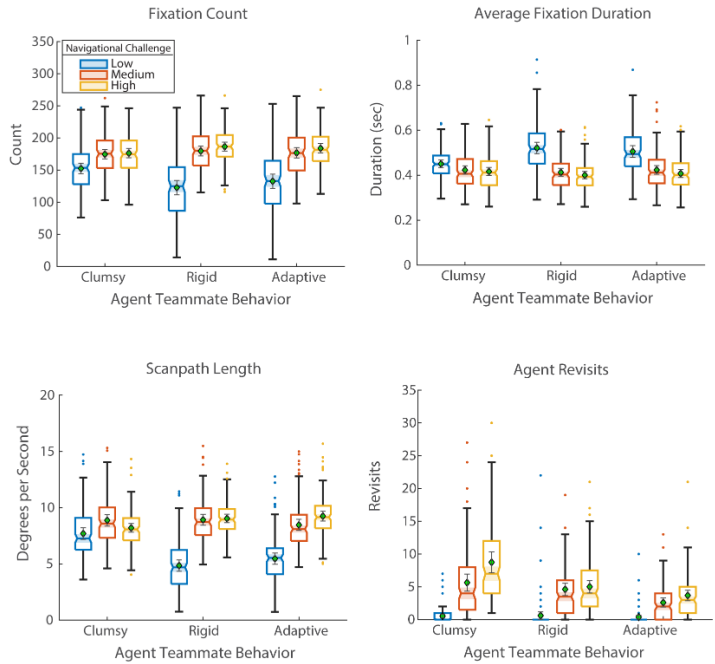

*Figure 7. Fixation count, fixation duration, scanpath length, and agent revisits for each agent, environment interaction.*

Overall, game play with the clumsy agent resulted in significantly more (clumsy-adaptive = 3.3, $t(134) = 2.67$, p = .03; clumsy-rigid = 4.82, $t(134) = 3.87$, $p < .001$, rigid-adaptive = not significant p = 0.7) and shorter (clumsy-adaptive = -.02, $t(134) = -4.77$, $p < .001$; clumsy-rigid = -0.01, $t(134) = -4.52$, $p < .001$, rigid-adaptive = not significant, p = 1) fixations and larger scanpaths (clumsy-adaptive = 0.53, $t(134) = 5.57$, $p < .001$; clumsy-rigid = 0.66, $t(134) = 6.86$, $p < .001$, rigid-adaptive = not significant, p = 0.6). The significant agent by environment interaction for fixation count and duration and scanpath length was driven primarily by environment for the rigid and adaptive agents (Fig 7). For these agents in the low nagivational complexity environment, eye fixations were fewer and longer, with smaller scanpath length relative to medium and high environments, suggesting less of a need to coordinate with the agent and more time for the human to focus on their own behavior. In the medium and high environments, eye fixations were more frequent and shorter with larger scanpath length. This suggests that higher navigational challenge environments significantly increase information processing demands, impacting how teammates monitor performance and coordinate tasks. The clumsy agent may have attracted more fixations relative to the other agents in the low complexity environment because of its unpredictable behavior.

Teammate performance monitoring as measured with revisits on the agent, indicated the adaptive agent was monitored the least followed by the rigid and clumsy agents (adaptive-rigid = -1.19, $t(134) = -4.24$, $p < .001$; adaptive-clumsy = -2.75, $t(134) = -9.82$, $p < .001$; rigid-clumsy = -1.56, $t(134) = -5.58$, $p < .001$). The reduced revisits to the adaptive agent, when compared to the clumsy and rigid agents, suggests less of a need to monitor and adjust to its actions leading to more efficient collaboration and team performance.

Higher navigational complexity resulted in more agent monitoring. Most revisits were found in the high complexity environment followed by medium and low, with all complexities significantly different from each other (high-low = 5.3, $t(134) = 16.11$, $p < .001$; medium-low = 3.77, $t(134) = 11.48$, $p < .001$; high-medium = 1.52, $t(134) = 4.63$, $p < .001$). This finding reflects a greater need to monitor and coordinate with the agent teammate as navigation becomes more challenging.

The significant interaction suggests that the increase in revisits as a function of navigational complexity varies depending on the agent. Differences in revisits between navigational complexities were largest for the clumsy agent and smallest for the adaptive agent. This suggests that less monitoring of the adaptive agent is required in environments with high navigational challenge when compared to the rigid and clumsy agents indicating better performance and predictability in complex environments.

*3.3.2. Gaze Allocation in Environment*

Differences in processing between items in the environment were analyzed by quantifying the proportion of eye gaze allocated to them. The proportion of gaze allocation to an element (i.e. agent, floor, soup, counter, onion and human) was separately analyzed due to their importance in task coordination and strategy. Each model revealed significant main effects and interactions ($p <= .004$, Supplement A.6). Figure 8 provides a breakdown of gaze allocation across items.

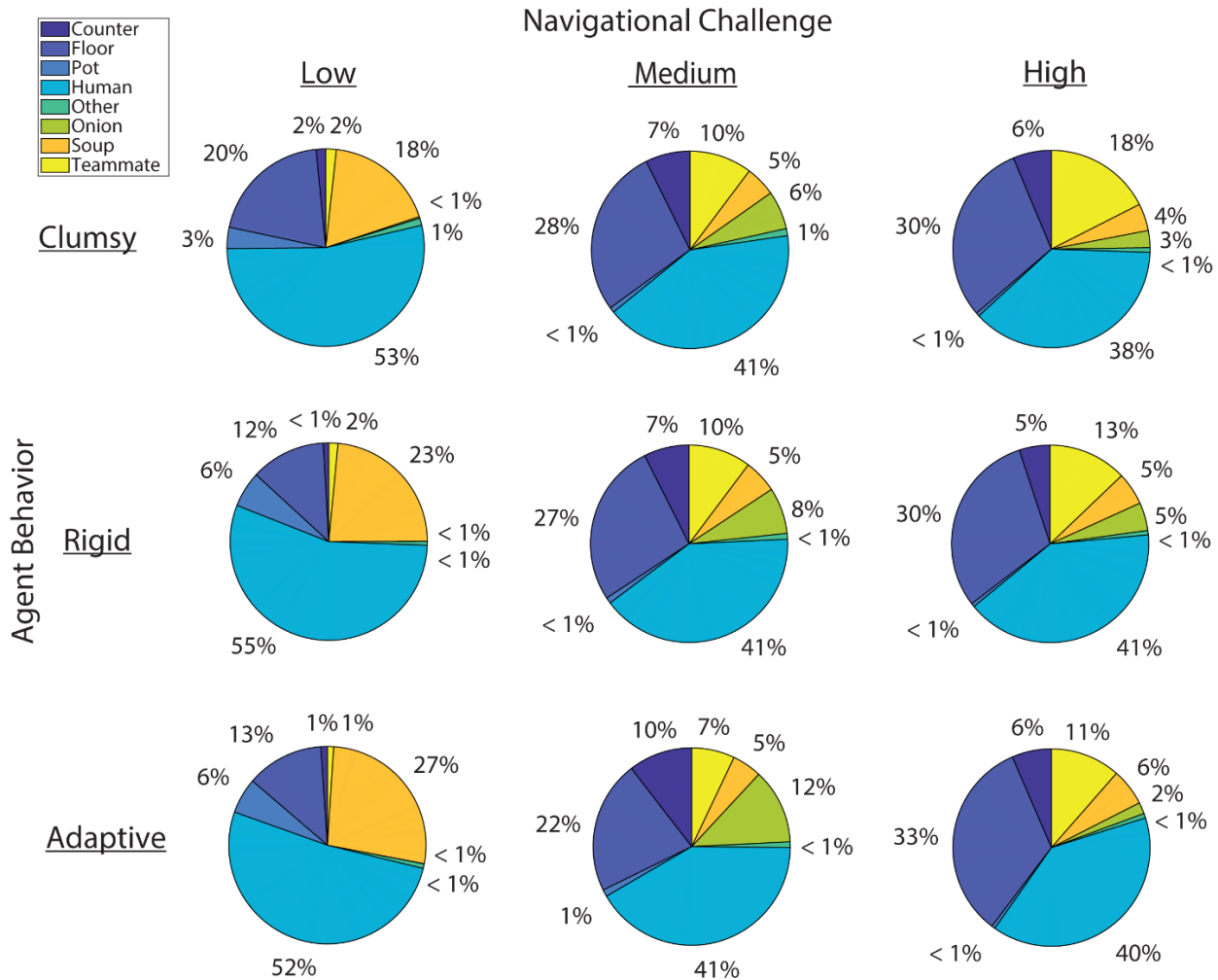

*Figure 8. Proportion of gaze samples allocated to items in the environment as a function of agent behavior and navigational challenge. Note. Percentages are rounded. 'Other' refers to gaze samples labeled as not a number (NaN), outside the game environment, dish dispenser, serving area, dish, and onion dispenser.*

Overall, the highest proportion of gaze was allocated to the human player. As navigational complexity increased, gaze on the human player decreased with increased gaze allocated to the teammate and the floor. This shift in gaze behavior was necessary due to the increasing need to coordinate movements with the teammate and plan the human player's own movements, reflected in increased gaze to the floor.

In the low navigational challenge environment, where the human player worked mostly independently, more gaze was allocated to the human player compared to the other environments. Additionally, more gaze was devoted to the soup in this environment, to monitor its completion, and to the floor, for navigation planning. Predictably, very little gaze was directed to the teammate in this environment due to the minimal need for team coordination.

In the medium environment, an interesting set of teaming behaviors emerged with a change in focus on different elements. As the agent became more capable, from clumsy to adaptive, more gaze was devoted to the onion and the counter and less to the floor. This reflects a strategy shift where passing the onions across the counter improves navigation efficiency but requires more sophisticated teaming coordination, which the adaptive agent exhibits best. There was a reduction of gaze on the teammate with the adaptive agent, presumably due to the more efficient and predictable strategy of focusing on the onions and the counter.

Within the high navigational environment, there was a clear reduction in gaze allocated to the teammate with the most gaze on clumsy agent, followed by the rigid agent, and the least on adaptive agent. This shift in attention reflects the competence of the agent and the reduced need to monitor and repeatedly adapt to its behavior. Additional gaze was then directed to the floor and the human for better navigational planning, movement, and soup delivery.

3.4. *Models of Relationships between Measures*

We tested two linear mixed effects models in R (using the 'lmer' and 'lme4' packages) to evaluate the role of eye tracking metrics in predicting agent trust and teaming behavior. Reported t-tests use Satterthwaites's method. In the trust model we aimed to determine whether the proportion of eye gaze on the agent teammate could predict agent trust, from post-trial trust ratings, beyond the effects of the navigational complexity and agent behavior. In the team contribution model, we examined if the contribution of the human teammate could be predicted by eye metrics associated with teaming and cognitive state beyond navigational challenge and agent behavior. Specifically, our predictors in the team contribution model included double-checking behavior (i.e. revisits), scanpath behavior (scanpath length per second), task engagement (blink count and duration), trust (subjective rating), and human gaze allocation (proportion of gaze on human player). Each model included a random effect for participant (see Table 1 for predictor variable definitions).

> **Trust Model**
>
> Trust ~ 1 + navigational challenge + agent behavior + proportion gaze on agent + (1|participant)
>
> **Team Contribution Model**
>
> Human contribution ~ 1 + navigational challenge + agent behavior + fixation revisits on agent + scanpath length + total blinks + blink duration + trust + proportion gaze on human + (1|participant)

In the trust model, no outliers surpassing a Cooks Distance of .02 were identified. The strongest fixed effects correlation observed was -.78 between the high navigational challenge environment and proportion of gaze on the agent. In the team contribution model, no outliers surpassing a Cooks Distance of .04 were found. The strongest fixed effects correlation observed was .79 between the clumsy agent and trust. In both models, visual inspection of scatterplots of residuals against fitted values showed no systematic deviations from zero, indicating homoscedasticity. Additionally, the histogram of residuals showed an approximately normal distribution for both models.

Analysis of the trust model (Table 2) revealed significant effects of each predictor, indicating that the proportion of eye gaze allocated to the agent was a significant predictor of trust over and above navigational challenge and teammate behavior. The inverse relationship shows that as gaze on the agent decreased, trust in the agent increased.

*Table 2. Output from linear mixed effects model predicting agent trust as a function of navigational challenge, agent teammate behavior and proportion of eye gaze on the agent.*

| Fixed Effects | Estimate | Std. Error | df | t | p |
|---|---|---|---|---|---|
| Intercept | 6.21 | 0.09 | 302.11 | 68.03 | <.001 |
| Nav Chal - High | 0.11 | 0.13 | 1213.35 | 0.86 | 0.39 |
| Nav Chal - Medium | -0.73 | 0.10 | 1196.91 | -6.99 | <.001 |
| Agent - Clumsy | -3.85 | 0.09 | 1164.95 | -44.21 | <.001 |
| Agent - Rigid | -0.50 | 0.08 | 1155.29 | -5.93 | <.001 |
| Gaze on Agent | -0.04 | 0.01 | 1214.81 | -4.53 | <.001 |

Note. Nav Chal indicates navigational challenge. Std. Error indicates standard error. df indicates degrees of freedom. t-tests use Satterthwaites's method

Analysis of the team contribution model indicated that certain eye metrics were more predictive than others (Table 3). Significant predictors included fixation revisits on the agent, scanpath length, blink count and trust. Each of these predictors decreased as the human contributed more to task completion. Fewer revisits on the agent suggests the human spent less time checking the agent and more time on completing subtasks, which was mirrored by decreased trust. The more the human contributed to the teaming behavior the less they trusted the agent. Increased human contribution to the team was also related to a decrease in scanpath length, suggesting a shift from global to local processing. This shift indicates increased processing within the immediate field of view rather than orienting to information outside of it. Eye blinks were significantly reduced with increased human team contribution implying greater potential for perceptual intake and increased task engagement during these periods. Eye blink duration and the proportion of eye gaze on the human player were not significant predictors in the model.

*Table 3. Output from linear mixed-effects model predicting human task contribution from navigational challenge, agent teammate behavior, fixation revisits on agent, scanpath frequency, blink duration, total blinks, trust, and proportion of gaze on the human player.*

| Fixed Effects | Estimate | Std. Error | df | t | p |
|---|---|---|---|---|---|
| Intercept | 0.352 | 0.0442 | 767.49 | 7.97 | <.001 |
| Nav Chal - High | -0.283 | 0.0140 | 1145.93 | -20.26 | <.001 |
| Nav Chal - Medium | -0.096 | 0.0133 | 1148.93 | -7.19 | <.001 |
| Agent - Clumsy | 0.547 | 0.0171 | 1128.15 | 32.06 | <.001 |
| Agent - Rigid | 0.161 | 0.0103 | 1083.97 | 15.56 | <.001 |
| Fixation Revisits on Agent | -0.002 | 0.0011 | 1141.74 | -2.17 | 0.030 |
| Scanpath Length | -0.026 | 0.0026 | 843.07 | -9.77 | <.001 |
| Blink Duration | 0.085 | 0.0539 | 1136.03 | 1.58 | 0.115 |
| Blink Count | -0.002 | 0.0007 | 457.86 | -2.87 | 0.004 |
| Trust | -0.015 | 0.0035 | 1148.67 | -4.13 | <.001 |
| Gaze on Human | 0.001 | 0.0005 | 958.90 | 1.38 | 0.167 |

Note. Nav Chal indicates navigational challenge. Std. Error indicates standard error. df indicates degrees of freedom. t-tests use Satterthwaites's method

## 4. Discussion

### 4.1. Summary

The goal of this study was to investigate how autonomous agents, trained with different strategies, work with humans in increasingly complex environments and how these interactions affect teaming, trust, and eye-gaze measures. While prior studies have examined trust and team performance in automation and human-agent interaction, our study is among the first to use eye-tracking metrics such as revisits, gaze allocation, and scanpath length to predict not just trust but human contribution to teaming behavior in a cooperative, real-time task with adaptive agents. We found that agent revisits increased with environmental complexity but decreased with agent adaptability, an interaction that offers a completely novel behavioral signature of teammate performance monitoring. Additionally, the inverse relationship between gaze on the agent and trust, while aligned with some prior literature (Patton & and Wickens, 2024), had not been demonstrated in this specific task structure or used in a predictive model of human-agent contribution. Furthermore, studies with true autonomy and true human-autonomy teaming remain rare (O'Neill et al., 2022b) with many studies simulating the autonomy (Riek, 2012) or instead using the classic automation dyadic paradigm that is not interdependent. This study uses a real AI teaming model to improve AI agents and human-autonomy teaming that exhibit both agency and interdependence. Lastly, a criticism of automation and human-autonomy teaming work is that researchers focus a lot on taskwork, not teamwork (McNeese et al., 2023). Our work is novel and contributes in that it does focus on teamwork both from the AI modeling side as well as from the setup between an AI and a human. These results therefore show unique insight into what true human-AI collaborative work looks like, how it can be assessed and further improved and that is a major contribution. We believe these findings extend the current literature by identifying practical, unobtrusive indicators that could be used to inform adaptive AI behavior in future HAT systems.

*4.2. Human-Autonomy Teaming with Adaptive AI Agents*

The adaptive AI teammate performed better and was perceived as more trustworthy, collaborative, and fluent than the other agents, replicating previous results with this platform and other environments (Aroca-Ouellette et al., 2023). Using hierarchical reinforcement learning, this agent better matched the task execution in the Overcooked AI environment, aligning with the game's task hierarchy and resulting in improved performance and predictability. Human-autonomy teaming improved primarily through the enhanced capabilities of the autonomous agent, with evidence showing that human performance increased as the agent's performance improved. When the agent is adaptive and anticipates human behavior, it not only improved its performance output but also improves teaming behavior. However, the largest variability was observed in the medium navigational environment indicating that not all humans could anticipate the agent's behavior. Some participants may therefore benefit from team building with an AI, a strategy that has proved effective in enhancing performance and collaboration (Walliser et al., 2019). Future research should focus on further enhancing teamwork between humans and autonomous agents by refining team processes and structures as well as optimizing the workspace in which they interact (Tung et al., 2024; Walliser et al., 2019).

Modelling agents based on meaningful tasks enhances teamwork by providing a clearer understanding of their behavior. This may fundamentally change how we approach HATs by shifting the focus from communication and transparency to a more personalized understanding of context and task structure (Begerowski et al., 2023; Mangin et al., 2022). Rather than designing agents for performance optimization or human-specific needs like mental model sharing, more focus should be on creating agents that excel in teamwork (McNeese et al., 2023). Designing for teamwork is more transferrable and foundational than many performance metrics.

To further improve the agent, other measures and methods may be needed to improve teamwork. For example, while Overcooked-AI serves as a good abstraction, the original Overcooked game provides more challenges and a higher level of complexity (Bishop et al., 2020). In human team play of this game, improved performance requires discussion of coordinated strategies, task allocation and motivation. Aside from communication, additional inputs may be needed to help the agent anticipate the human teammate's behavior. This becomes even more important when the team scales to multiple agents and humans (Hertz et al., 2019; Walliser et al., 2023).

Understanding and transparency of the teammate are important for effective collaboration (Endsley, 2023; Walliser et al., 2023). In this study, the adaptive agent excelled in this area by having traceable goals and tasks, allowing it to share its intent more effectively. Some have already demonstrated this in the context of Overcooked AI and robot collaboration showing that increased transparency through communicating intent and task planning, leads to better collaboration with the agent and decreases cognitive load (Berberian et al., 2023; Le Guillou et al., 2023; Roncone et al., 2017). Foundational models, like large language models (LLMs) could prove useful in this context. Combining transparency or trust cues (E. J. de Visser et al., 2014) with an agent trained using a task hierarchy can further facilitate helpful human-autonomy teaming.

### 4.3. *Eye Metrics as Indicators of Human-Autonomy Teamwork*

Eye-gaze metrics have proven valuable for assessing team task engagement and detecting levels of trust within human-autonomy teaming . Previous research has identified a negative correlation between the degree of monitoring in automation and trust, where increased monitoring often signifies distrust (Patton & Wickens, 2024). This study reinforces those findings, demonstrating the proportion of gaze on the agent teammate was a significant predictor of trust. Eye tracking captures the additional monitoring required to oversee unreliable teammates, with studies noting increased gaze allocation to erroneous actions performed by automated teammates (Wachowiak et al., 2022). Our findings revealed that participants frequently looked at their own avatars, consistent with previous research indicating this behavior is common during normal workflow (Wachowiak et al., 2022).

Interpreting gaze allocation metrics, however, requires caution. The proportion of gaze allocated to an item is often equated with the 'amount' of attention given to that item. While overt gaze shifts orient attention to visual stimuli, information may still be selected and prioritized using covert attention shifts, i.e. shifting attention with little to no eye movement (Bisley, 2011; Handy & Khoe, 2005; Henderson et al., 1989; Posner, 2016). In our experiment, participants might have tracked the agent using covert attention even though their gaze position was on a different item, or alternatively, processing environmental elements when gaze was on the agent. Thus, the proportion of gaze allocated to specific items should be interpreted in this broader context.

The structured nature of our experimental environments allowed us to infer team strategies based on agent revisits, underscoring the utility of this metric in understanding team dynamics. Specifically, our data revealed that revisits on the autonomous teammate increased with navigational complexity but decreased with agent versatility. This pattern suggests that in more complex environments, participants felt a greater need to monitor their teammates closely, reflecting higher load and coordination demands. Conversely, the decrease in revisits with more versatile (adaptive) agents indicates that these agents were more predictable and reliable, reducing the need for constant monitoring. It is important to note that fixation revisits are distinctly different from gaze allocation. In this context, agent revisits refer to double-checking or backtracking behavior, where the agent is immediately re-fixated after fixating on a different item. This metric does not represent the total time spent looking at the agent, as measured by the proportion of eye gaze metric.

Practically speaking, eye-tracking data can be used as input for AI teammates to adapt dynamically to human users, aligning with the concept of adaptive automation (E. A. Byrne & Parasuraman, 1996; Feigh et al., 2012; Inagaki, 2003; Kaber & Endsley, 2004; Parasuraman et al., 2009; Scerbo, 2018). Prior studies have demonstrated that eye-gaze indicators of human errors can be used as triggers for automation assistance (Gartenberg et al., 2013; Ratwani et al., 2008; Ratwani & Trafton, 2011). Such input helps AI agents anticipate human behavior and re-prioritize tasks accordingly. Recent work by our group supports this approach, showing that models combining eye-tracking metrics with other human inputs better predict proficiency, trust, and intention than models using either input alone (Hulle et al., 2024).

This work demonstrates that eye-tracking metrics offer unobtrusive indicators of human-autonomy teaming and performance. **Our findings suggest that gaze features could serve as triggers for adaptive agent behavior**. For example, increased agent revisits may prompt the agent to simplify its actions or increase transparency (Mercado et al., 2016), while reduced blink count and scanpath length may signal high engagement, suggesting the agent could scale back its involvement. Conversely, broader scanpaths or gaze shifts away from the agent may reflect planning or declining situation awareness, prompting increased agent support. Together, these gaze-informed signals can inform dynamic adjustments in agent behavior to support key teaming constructs such as trust, workload, awareness, and planning (Bolstad & Endsley, 2000; Demir et al., 2017; Gorman et al., 2017).

*4.4. Limitations*

Given our display size and game space, task relevant information was confined to a small portion of the available visual field. However, in many real-world situations, we naturally incorporate information from beyond this restricted area through combined eye and head movements. Prior research suggests that behavior changes when head movements are involved in acquiring peripheral, goal-relevant information (David et al., 2020; Draschkow et al., 2021; Solman & Kingstone, 2014) Specifically, we tend to rely more on memory in situations that require large eye/head movements (Draschkow et al., 2021). Therefore, future research should evaluate how eye gaze metrics related to teammate performance monitoring and team coordination change when head movements are also necessary for monitoring and coordinating teaming behavior.

The observed teaming behavior in this context may have been specific to the environments we chose to work with. However, there are many alternative configurations that could change teaming behavior in this task. For example, how may teaming behavior change with increased task complexity (e.g. multiple ingredients, player specific roles) or team composition (multiple humans or agents in the same environment)?

Eye tracking measures may indicate multiple human-autonomy teaming processes. For example, revisits might suggest a lack of trust in the agent, but they could also indicate strategic adaptation to the teammate's behavior or support for back-up behavior (Salas et al., 2005). Additionally, many of our measures are task-specific and require context to interpret the eye-tracking behavior accurately, such as knowing exactly what participants are looking at. To generalize these findings to other tasks, the information and conditions in the environment are crucial. Other factors that may influence eye-tracking measures include how the agent is represented (e.g., avatar, robot, more human-like interfaces) and the task environment (E. J. de Visser et al., 2016).

Lastly, the human teaming constructs of adaptability and performance monitoring are richer in their full realization and were only partially observed in this context. For example, the construct back-up behavior in its full definition is *"the ability to anticipate other team member's needs through accurate knowledge about their responsibilities. This includes the ability to shift workload among members to achieve balance during high periods of workload or pressure"*

(Salas et al., 2005, p. 560, Table 1). While we observed some anticipation by the human participants of the AI's responsibilities, shifting workload to achieve balance would be a more advanced teaming behavior that was not observed in the current study. Future human-autonomy teaming may demonstrate this behavior in different type of environments requiring more interdependence and coordination and with AI agents that can communicate about task and teamwork

**Conclusion**

This study demonstrates that AI agents designed with hierarchical reinforcement learning can significantly improve human-autonomy teaming by enhancing coordination, balancing task contributions, and fostering higher trust. Through dynamic teaming scenarios in Overcooked AI, we identified specific eye-tracking metrics such as fixation revisits, blink count, scanpath length, and gaze allocation that reliably reflect changes in trust and predict human contributions to team performance. These results advance our understanding of how gaze behavior can serve as an unobtrusive indicator of teaming constructs related to trust calibration, workload, and situation awareness. Importantly, this work is among the first to show that eye metrics can be used not only to assess but potentially to inform adaptive agent behavior. Together, these findings support the use of gaze-informed models as a foundation for developing more responsive, trustworthy, and effective human-autonomy teams.

**Declaration of generative AI and AI-assisted technologies in the writing process:**
The authors used ChatGPT4o to concisely refine some sentences. After using this tool, the authors reviewed and edited the content as needed and take full responsibility for the content of the published article.

**Data availability**

Data will be made available upon request.

Supplementary material

Supplement A.1 Percentage data loss for each participant (color) and trial (individual points). Top: Participants having no trials with greater than 30% data loss. Bottom: Participants having at least one trial with greater than 30% data loss.

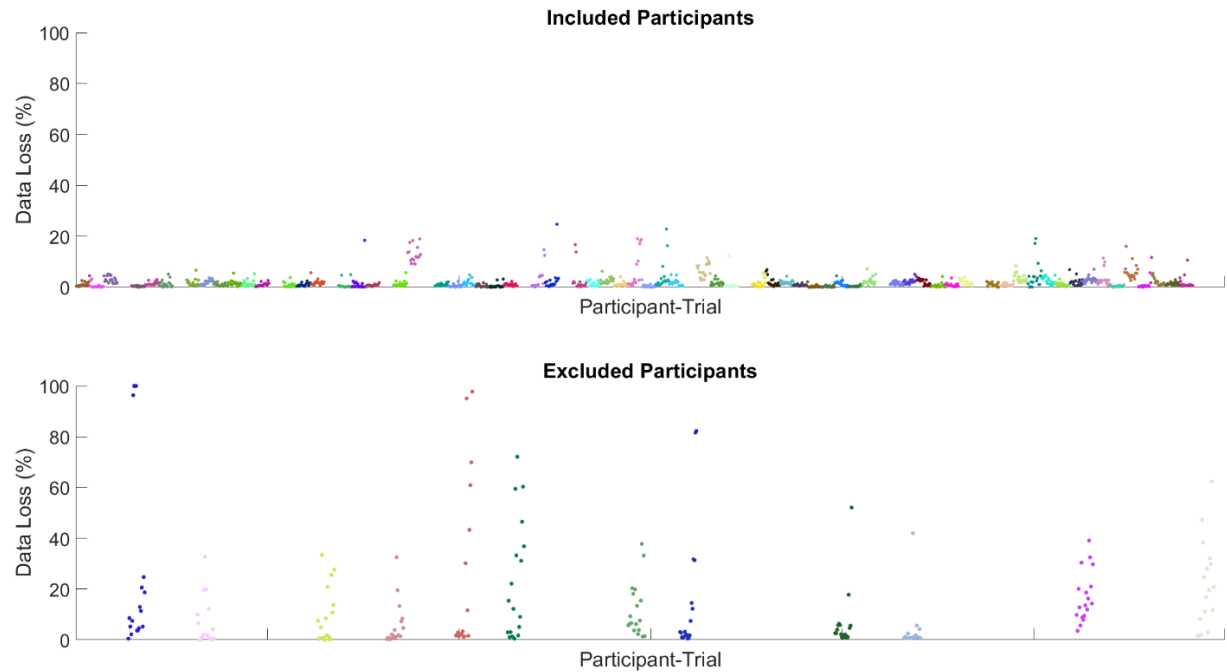

Supplement A.2

*ANOVA output for collision and score analyses.*

| DV | IV | $df_{Num}$ | $df_{Den}$ | Epsilon | $SS_{Num}$ | $SS_{Den}$ | F | p | $\eta^2$ |
|---|---|---|---|---|---|---|---|---|---|
| Collisions | Navigational Challenge | 1 | 67 | | 10260.09 | 5085.66 | 135.2 | <.001 | 0.67 |
| | Agent Behavior | 1.59 | 107 | 0.79 | 20654.9 | 18445.69 | 75.02 | <.001 | 0.53 |
| | Navigational Challenge * Agent Behavior | 1.66 | 111 | 0.83 | 736.79 | 9674.46 | 5.1 | 0.011 | 0.07 |
| Score | Navigational Challenge | 1.11 | 74.3 | 0.55 | 2246906 | 182160.8 | 826.4 | <.001 | 0.93 |
| | Agent Behavior | 1.86 | 125 | 0.93 | 1379336 | 75530.39 | 1224 | <.001 | 0.95 |
| | Navigational Challenge * Agent Behavior | 2.23 | 149 | 0.56 | 136504.9 | 128161.8 | 71.36 | <.001 | 0.52 |

Note. DV indicates dependent variable and IV indicates independent variable. dfNum indicates degrees of freedom numerator. dfDen indicates degrees of freedom denominator. Epsilon indicates Greenhouse-Geisser multiplier for degrees of freedom, p-values and degrees of freedom in the table incorporate this correction. SSNum indicates sum of squares numerator. SSDen indicates sum of squares denominator. η2 indicates partial eta-squared.

# Supplement A.3

*ANOVA output for total number of subtasks completed by each team member and the contribution of subtasks toward the task goal.*

| DV | IV | $df_{Num}$ | $df_{Den}$ | Epsilon | $SS_{Num}$ | $SS_{Den}$ | F | p | $\eta^2$ |
|---|---|---|---|---|---|---|---|---|---|
| Subtasks | Team member | 1 | 67 | | 105599.8 | 20665.39 | 342.4 | <.001 | 0.92 |
| | Agent Behavior | 1.94 | 130 | 0.97 | 128657.4 | 10486.53 | 822 | <.001 | 0.91 |
| | Navigational Challenge | 1.47 | 98.4 | 0.73 | 164613.7 | 15970.99 | 690.6 | <.001 | 0.84 |
| | Navigational Challenge * Agent Behavior | 2.91 | 195 | 0.73 | 6531.33 | 14874.12 | 29.42 | <.001 | 0.31 |
| | Agent Behavior * Team member | 1.86 | 125 | 0.93 | 64599.36 | 4235.86 | 1022 | <.001 | 0.94 |
| | Navigational Challenge * Team member | 1.9 | 128 | 0.95 | 127862.4 | 8872.68 | 965.5 | <.001 | 0.94 |
| | Navigational Challenge * Agent Behavior * Team member | 3.24 | 217 | 0.81 | 4311.29 | 7031.49 | 41.08 | <.001 | 0.38 |
| Contribution | Navigational Challenge | 1.58 | 106 | 0.79 | 13.95 | 1.58 | 589.9 | <.001 | 0.9 |
| | Agent Behavior | 1.99 | 133 | 0.99 | 36.8 | 1.59 | 1551 | <.001 | 0.96 |
| | Navigational Challenge * Agent Behavior | 2.99 | 200 | 0.75 | 1.41 | 2.6 | 36.49 | <.001 | 0.35 |

Note. DV indicates dependent variable and IV indicates independent variable. dfNum indicates degrees of freedom numerator. dfDen indicates degrees of freedom denominator. Epsilon indicates Greenhouse-Geisser multiplier for degrees of freedom, p-values and degrees of freedom in the table incorporate this correction. SSNum indicates sum of squares numerator. SSDen indicates sum of squares denominator. η2 indicates partial eta-squared.

Supplement A.4

*ANOVA output for subjective agent ratings*

| Post-trial Statement | IV | $df_{Num}$ | $df_{Den}$ | Epsilon | $SS_{Num}$ | $SS_{Den}$ | F | p | $\eta^2$ |
|---|---|---|---|---|---|---|---|---|---|
| *The human-agent team worked fluently together* | Navigational Challenge | 1.56 | 104.34 | 0.78 | 144.35 | 129.37 | 74.76 | <.001 | 0.53 |
|  | Agent Behavior | 1.85 | 123.91 | 0.92 | 1842.91 | 131.81 | 936.73 | <.001 | 0.93 |
|  | Navigational Challenge * Agent Behavior | 3.29 | 220.73 | 0.82 | 81.44 | 212.34 | 25.7 | <.001 | 0.28 |
| *I was the most important team member* | Navigational Challenge | 1.43 | 95.91 | 0.72 | 21.4 | 98.93 | 14.49 | <.001 | 0.18 |
|  | Agent Behavior | 1.88 | 125.86 | 0.94 | 891.68 | 141.65 | 421.77 | <.001 | 0.86 |
|  | Navigational Challenge * Agent Behavior | 3.53 | 236.37 | 0.88 | 25.95 | 142.22 | 12.22 | <.001 | 0.15 |
| *I trusted the agent to do the right thing* | Navigational Challenge | 1.83 | 122.74 | 0.92 | 108.62 | 112.82 | 64.5 | <.001 | 0.49 |
|  | Agent Behavior | 1.79 | 119.96 | 0.9 | 1881.41 | 152.03 | 829.14 | <.001 | 0.93 |
|  | Navigational Challenge * Agent Behavior | 3.29 | 220.74 | 0.82 | 70.7 | 197.52 | 23.98 | <.001 | 0.26 |
| *I understood what the agent was trying to accomplish* | Navigational Challenge | 1.87 | 125.23 | 0.93 | 79.71 | 84.79 | 62.99 | <.001 | 0.48 |
|  | Agent Behavior | 1.69 | 112.96 | 0.84 | 2130.08 | 154.42 | 924.23 | <.001 | 0.93 |
|  | Navigational Challenge * Agent Behavior | 3.5 | 234.67 | 0.88 | 62.97 | 188.69 | 22.36 | <.001 | 0.25 |
| *The agent was cooperative* | Navigational Challenge | 1.7 | 113.75 | 0.85 | 111.05 | 129.06 | 57.65 | <.001 | 0.46 |
|  | Agent Behavior | 1.92 | 128.79 | 0.96 | 1938.4 | 161.55 | 803.92 | <.001 | 0.92 |
|  | Navigational Challenge * Agent Behavior | 3.24 | 217.35 | 0.81 | 79.5 | 224.05 | 23.77 | <.001 | 0.26 |

Note. IV indicates independent variable. dfNum indicates degrees of freedom numerator. dfDen indicates degrees of freedom denominator. Epsilon indicates Greenhouse-Geisser multiplier for degrees of freedom, p-values and degrees of freedom in the table incorporate this correction. SSNum indicates sum of squares numerator. SSDen indicates sum of squares denominator. η2 indicates partial eta-squared.

## Supplement A.5

*ANOVA output for fixation count, fixation duration, scanpath length, and agent revisits.*

| DV | IV | $df_{Num}$ | $df_{Den}$ | Epsilon | $SS_{Num}$ | $SS_{Den}$ | F | p | $\eta^2$ |
|---|---|---|---|---|---|---|---|---|---|
| Fixation Count | Navigational Challenge | 1.35 | 90.23 | 0.67 | 262933.2 | 55471 | 317.58 | <.001 | 0.83 |
| | Agent Behavior | 1.98 | 132.94 | 0.99 | 2478.1 | 21121 | 7.86 | 0.001 | 0.11 |
| | Navigational Challenge * Agent Behavior | 3.21 | 214.78 | 0.8 | 33302.85 | 39856 | 55.98 | <.001 | 0.46 |
| Fixation Duration | Navigational Challenge | 1.46 | 97.88 | 0.73 | 0.85 | 0.25 | 228.33 | <.001 | 0.77 |
| | Agent Behavior | 1.93 | 129.23 | 0.96 | 0.03 | 0.14 | 14.43 | <.001 | 0.18 |
| | Navigational Challenge * Agent Behavior | 3.18 | 213.03 | 0.79 | 0.17 | 0.3 | 38.36 | <.001 | 0.36 |
| Scanpath Length | Navigational Challenge | 1.86 | 124.59 | 0.93 | 1058.44 | 241.28 | 293.91 | <.001 | 0.81 |
| | Agent Behavior | 1.97 | 132.06 | 0.99 | 49.51 | 124.65 | 26.61 | <.001 | 0.28 |
| | Navigational Challenge * Agent Behavior | 3.39 | 227.29 | 0.85 | 307.15 | 226.53 | 90.84 | <.001 | 0.58 |
| Agent Revisits | Navigational Challenge | 1.66 | 111.34 | 0.83 | 3036.25 | 1478.7 | 137.57 | <.001 | 0.67 |
| | Agent Behavior | 1.62 | 108.35 | 0.81 | 774.78 | 1070.7 | 48.48 | <.001 | 0.42 |
| | Navigational Challenge * Agent Behavior | 2.95 | 197.75 | 0.74 | 487.69 | 1722.9 | 18.97 | <.001 | 0.22 |

Note. DV indicates dependent variable and IV indicates independent variable. dfNum indicates degrees of freedom numerator. dfDen

# Supplement A.6

*ANOVA output for proportion of gaze samples on items in the environment.*

| Gaze on: | IV | $df_{Num}$ | $df_{Den}$ | Epsilon | $SS_{Num}$ | $SS_{Den}$ | F | p | $\eta^2$ |
|---|---|---|---|---|---|---|---|---|---|
| Agent | Navigational Challenge | 1.85 | 123.65 | 0.92 | 14355.2 | 1875.05 | 512.94 | <.001 | 0.88 |
|  | Agent Behavior | 1.7 | 113.61 | 0.85 | 933.18 | 972.64 | 64.28 | <.001 | 0.49 |
|  | Navigational Challenge * Agent Behavior | 3.41 | 228.14 | 0.85 | 610.41 | 1672.12 | 24.46 | <.001 | 0.27 |
| Floor | Navigational Challenge | 1.92 | 128.94 | 0.96 | 24469.7 | 4106.61 | 399.23 | <.001 | 0.86 |
|  | Agent Behavior | 1.89 | 126.67 | 0.95 | 914.66 | 1758.8 | 34.84 | <.001 | 0.34 |
|  | Navigational Challenge * Agent Behavior | 3.61 | 242.04 | 0.9 | 2609.01 | 2997.9 | 58.31 | <.001 | 0.47 |
| Soup | Navigational Challenge | 1.09 | 73.15 | 0.55 | 36721 | 18146.9 | 135.58 | <.001 | 0.67 |
|  | Agent Behavior | 1.88 | 126.21 | 0.94 | 1198.1 | 1794.32 | 44.74 | <.001 | 0.4 |
|  | Navigational Challenge * Agent Behavior | 2.5 | 167.51 | 0.63 | 1454.67 | 3283.64 | 29.68 | <.001 | 0.31 |
| Counter | Navigational Challenge | 1.94 | 130.05 | 0.97 | 4758.59 | 843.56 | 377.95 | <.001 | 0.85 |
|  | Agent Behavior | 1.95 | 130.69 | 0.98 | 225.28 | 307.97 | 49.01 | <.001 | 0.42 |
|  | Navigational Challenge * Agent Behavior | 2.98 | 199.42 | 0.74 | 254.15 | 745.91 | 22.83 | <.001 | 0.25 |
| Onion | Navigational Challenge | 1.25 | 83.78 | 0.63 | 6720.12 | 2390.67 | 188.34 | <.001 | 0.74 |
|  | Agent Behavior | 1.95 | 130.83 | 0.98 | 246.99 | 630.26 | 26.26 | <.001 | 0.28 |
|  | Navigational Challenge * Agent Behavior | 2.55 | 170.95 | 0.64 | 1212.18 | 1497.82 | 54.22 | <.001 | 0.45 |
| Human | Navigational Challenge | 1.35 | 90.35 | 0.67 | 20585.6 | 18691.6 | 73.79 | <.001 | 0.52 |
|  | Agent Behavior | 1.99 | 133.66 | 1 | 338.51 | 3995.91 | 5.68 | 0.004 | 0.08 |
|  | Navigational Challenge * Agent Behavior | 2.8 | 187.34 | 0.7 | 536.3 | 6713.57 | 5.35 | 0.002 | 0.07 |

Note. dfNum indicates degrees of freedom numerator. dfDen indicates degrees of freedom denominator. Epsilon indicates Greenhouse-Geisser multiplier for degrees of freedom, p-values and degrees of freedom in the table incorporate this correction. SSNum indicates sum of squares numerator. SSDen indicates sum of squares denominator. η2 indicates partial eta-squared.

Supplement A.7
General gaming experience and specific experience with Overcooked

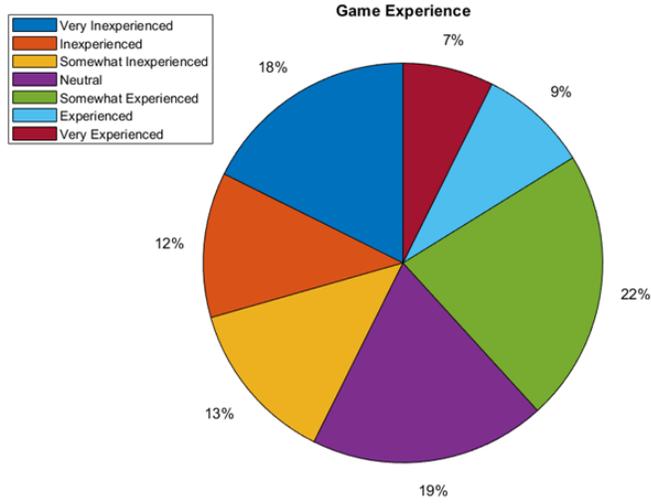
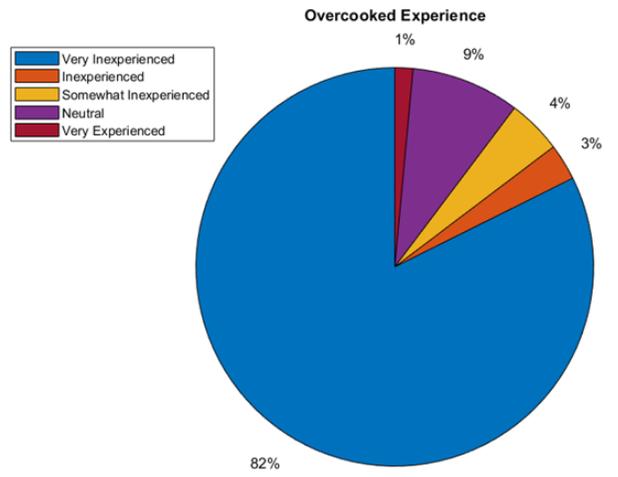